\documentclass[]{aastex631}

\begin{document}

\title{Modeling the Solar Transition Region: Effects of Spatial Resolution on the Atmospheric Structure, Emission and Non-Equilibrium Ionization}

\author[0000-0002-1043-9944]{Takuma Matsumoto}
\affiliation{Centre for Integrated Data Science, Institute for Space-Earth Environmental Research, Nagoya University, Furocho, Chikusa-ku, Nagoya, Aichi 464-8601, Japan}
\affiliation{National Astronomical Observatory of Japan, 2-21-1 Osawa, Mitaka, Tokyo 181-8588, Japan}

\begin{abstract}
The solar transition region (TR) is a narrow interface between the chromosphere and corona, where emitted radiation contains critical information pertinent to coronal heating processes. 
We conducted 2-dimensional radiation magnetohydrodynamics simulations using adaptive mesh refinement to spatially resolve the fine structure of the TR while simultaneously capturing the larger-scale dynamics originating from surface convection. 
The time evolution of ionization fractions for oxygen ions is computed alongside the simulations. 
A minimum grid size of 1.25 km is achieved in the TR, enabling adequate resolution of the upper TR (log$_{10}T \gtrsim$ 5), although the lower TR (log$_{10}T \lesssim$ 5) remains under-resolved. 
Doppler shifts and nonthermal widths synthesized from TR lines exhibit convergence with grid sizes as coarse as 40 km, though some discrepancies persist between our results and observed TR line properties. 
A notable enhancement in emission from \ion{O}{6} lines, converging at a grid size of 2.5 km, shows an intensity 1.2 times that expected under ionization equilibrium, attributable to shock interactions with the TR. 
While model refinements are still required, our ability to resolve the TR offers critical insights into TR line characteristics arising from non-equilibrium ionization states, advancing our understanding of the coronal heating problem.
\end{abstract}


\section{Introduction} \label{sec:intro}

The solar transition region (TR) represents a sharp boundary between the cool, dense chromosphere and the hot, tenuous corona. 
Although the TR is considerably thinner than the local pressure scale height, diagnosing this region is crucial for studying the coronal heating problem, as all energy and mass must ultimately pass through it. 
Recent observations indicate that energy and mass exchanges through the TR are highly dynamic and inhomogeneous processes, driven by the interaction between magnetic fields and surface convection zones \citep[see review by][ and references therein]{2017RAA....17..110T}. 
Therefore, it is essential to develop numerical models that capture both the global structure, extending from surface convection zones to the corona, and the thin, winding TR simultaneously.

The coronal density is known to be significantly underestimated in numerical simulations that do not fully resolve the thickness of the TR \citep{2013ApJ...770...12B}. 
The estimated thickness of the upper TR, where $T \sim 2 \times 10^5$ K, is approximately 50 km, with the lower TR being even thinner \citep{2006SoPh..234...41K}. 
Therefore, spatial resolutions on the order of a few kilometers or less are required, a stringent condition for multi-dimensional simulations. 
This requirement can be mitigated by artificially broadening the TR thickness \citep{2009ApJ...690..902L,2019ApJ...873L..22J,2021ApJ...917...65I}, allowing the coronal density to be accurately estimated even with coarse grid sizes.
Although the response of the corona to artificial heating has been tested \citep{2021A&A...654A...2J}, the broadening technique may introduce unintended consequences in coronal heating processes, such as the energy injection via Alfv\'{e}n waves \citep{2023MNRAS.526..499H}.
Furthermore, these methods do not ensure to spatially resolve the TR structure itself, which is critical to synthesize UV lines from the TR.

Resolving the TR thickness is also crucial for understanding the behavior of ions in non-equilibrium ionization (NEI).
Solving the advection equations for these ions within the narrow TR requires fine numerical grids, as coarse grid spacing can lead to significant numerical diffusion. 
This diffusion can introduce unphysical deviations from the equilibrium state. 
While multi-dimensional simulations incorporating NEI have been performed \citep{2013ApJ...767...43O, 2015ApJ...802....5O, 2016ApJ...817...46M, 2017ApJ...850..153N}, particular caution is necessary regarding the effects of under-resolved TR, especially for Li-like ions \citep{2024ApJ...964..107M}, which have narrow formation temperature ranges and exhibit anomalous emission \citep{2002A&A...385..968D}.

The aim of this study is to develop a numerical model capable of resolving the fine structure of the TR while simultaneously capturing the global dynamics from the surface convection zone. 
To achieve this, we will employ the adaptive mesh refinement (AMR) technique, which has been primarily applied in one-dimensional simulations \citep{2003A&A...401..699B}, to model the solar atmosphere. 
By resolving the TR with high precision, this approach aims to accurately reproduce both the coronal structure and ionization states, which might be misrepresented in coarser models. 
The model will further allow for a detailed examination of the interaction between the TR and convection-driven waves and flows, crucial for understanding energy and mass transport through the TR. 
Additionally, incorporating NEI effects will enable us to investigate the emission from ions that deviate from equilibrium in the TR, particularly within multi-dimensional simulations.
Ultimately, the goal is to improve our understanding of coronal heating processes and the role of the TR in facilitating energy and mass transfer between the solar atmosphere layers.

\section{Models and Assumptions} \label{sec:model}

{In this study, we have developed a completely new simulation code to model the solar atmosphere above the quiet Sun, extending from the upper convection to the corona.}
The model incorporates magnetohydrodynamics (MHD), radiative transfer, and the general equation of state to accurately represent the dynamics of the surface convection zone. 
The cooling of the corona, governed by optically thin radiation and thermal conduction, is balanced by an artificial heating term near the top boundary.
Additionally, the time-dependent ionization and recombination processes of oxygen ions are included to synthesize the (E)UV line intensities.

To realize all the above mentioned processes, we numerically solved the following basic equations:
\begin{eqnarray}
    \frac{\partial \rho}{\partial t} &+& \nabla \cdot \left( \rho \mathbf{v} \right) = 0, \\
    \frac{\partial \rho \mathbf{v}}{\partial t} &+& \nabla \cdot \left(
    \rho \mathbf{vv} + P_T - \mathbf{BB}
    \right) = \rho \mathbf{g}, \label{eq:eom}\\
    \frac{\partial \mathbf{B}}{\partial t} &+& \nabla \cdot \left(
    \mathbf{vB} - \mathbf{Bv} + \psi 
    \right) = 0, \label{eq:induction}\\
    \frac{\partial {\cal E}}{\partial t} &+& \nabla \cdot \left[
    \left( {\cal E} + P_T \right) \mathbf{v} -
    \mathbf{B} \left( \mathbf{v} \cdot \mathbf{B} \right)
    - \mathsf{\kappa} \nabla T
    \right] \nonumber \\
    &=& \rho \mathbf{v}\cdot \mathbf{g} + Q_{\rm rad} + Q_{\rm heat}, \label{eq:eoe} \\
    \frac{\partial \psi}{\partial t} &+& \nabla \cdot \left(
    c_h^2 \mathbf{B}
    \right) = -\frac{c_h^2}{c_p^2} \psi \label{eq:divbclean},
\end{eqnarray}
where $\rho, T, \mathbf{v},$ and $\mathbf{B}$ are, mass density, gas temperature, velocity, and magnetic field (normalized by $\sqrt{4\pi}$), respectively.
The total pressure, $P_{\rm T} = P_{\rm g} + B^2/2$ and the total energy, ${\cal E} = (\rho v^2 + B^2)/2 + e_{\rm int}$, where $P_{\rm g}$ and $e_{\rm int}$ are gas pressure and internal energy, respectively.
We assumed constant gravitational acceleration in the vertical direction, $g=-2.74\times10^4$ cm s$^{-2}$.
For thermal conduction, Spitzer-H\"{a}rm flux along magnetic field is adopted:
\begin{eqnarray}
    \kappa \nabla T = \kappa_0 T^{5/2} (\nabla T \cdot \mathbf{B}) \mathbf{B} / B^2,
\end{eqnarray}
where $\kappa_0=10^{-6}$ erg cm$^{-1}$ s$^{-1}$ K$^{-7/2}$.
To mitigate the underestimation of coronal density caused by limited grid spacing in the TR, we implemented the LTRAC method \citep{2021ApJ...917...65I}, enhancing thermal conductivity locally while adjusting the radiative cooling rate.
The LTRAC method was applied to all simulations, regardless of the maximum AMR level, to ensure consistency.
We adopted the tabulated equation of state provided by Modules for Experiments in Stellar Astrophysics (MESA) \citep{2011ApJS..192....3P}.
For the chemical fraction in the metal, we used the abundance in \cite{2009ARA&A..47..481A}.
%$(X,Y,Z)=(0.732,0.254,0.014)$.
To reduce the numerical magnetic monopoles, we employed the divergence cleaning method \citep{2002JCoPh.175..645D}.
This leads us to solve an additional equation for $\psi$ (Equation \ref{eq:divbclean}) and modify the induction equation (Equation \ref{eq:induction}).
The MHD equations are solved numerically using the HLLD scheme \citep{2005JCoPh.208..315M} with a second-order van Leer limiter, integrated using a predictor–corrector midpoint method.

To evaluate the cooling rate in optically thick regions, we solved radiative transfer equations assuming local thermal equilibrium:
\begin{eqnarray}
    \hat{n}\cdot \nabla I = \rho \kappa_{\rm R} \left( \frac{\sigma}{\pi} T^4 - I \right),
\end{eqnarray}
where $\hat{n},I,\kappa_{\rm R}$, and $\sigma$ are unit direction vector of a ray, specific intensity, Rosseland mean opacity, and Stefan–Boltzmann constant, respectively.
The radiative cooling rate is then derived with 
\begin{eqnarray}
    Q_{\rm thick} = 4 \pi \kappa_{\rm R} (J - B),  
\end{eqnarray}
where $J$ and $B$ are mean intensity and Planck function, respectively.
{To solve the radiative transfer equation, we employed the short characteristics method \citep{1988JQSRT..39...67K}, which is well-suited for parallel computing and applicable to unstructured meshes \citep{1999A&A...348..233B}.}
We adopted the opacity tables provided by MESA.

For optically thin cooling, we adopted combination of two different cooling functions \citep{2015ApJ...812L..30I}.
We used CHIANTI database \citep{2021ApJ...909...38D} with photospheric abundance for high temperature (T $\gtrsim$ 15,000 K), while we used the formulation from \citep{2012ApJ...751...75G} for the lower temperature.
Using the combined cooling function, $\Lambda (T)$, the cooling rate can be expressed as
\begin{eqnarray}
    Q_{\rm thin} = -n_{\rm e} n_{\rm H} \Lambda (T), 
\end{eqnarray}
where $n_{\rm e}$ and $n_{\rm H}$ are electron and hydrogen number density, respectively.

After calculating $Q_{\rm thick}$ and $Q_{\rm thin}$, the total radiative loss is derived by
\begin{eqnarray}
    Q_{\rm rad} &=& \xi Q_{\rm thin}  + (1 - \xi) Q_{\rm thick} , \\
    \xi &=& \exp (- P_{\rm g} / P_{*})
\end{eqnarray}
where $P_{*} = $ 10 erg cm$^{-3}$ in order to smoothly connect the cooling rates in two different regime. 

To sustain the high-temperature corona, we introduced an artificial heating term near the upper boundary, expressed as:
\begin{eqnarray}
    Q_{\rm heat} = \frac{\rho R_{\rm g}~{\rm max}(T_{\rm top} - T,0)}{\tau} \exp (-\frac{(z-z_{\rm top})^2}{2w^2}),
\end{eqnarray}
where $R_{\rm g}$ represents the gas constant.
The parameter $T_{\rm top}$ and $w$ is set to $10^6$ K and 400 km, with $\tau$ set to 10 numerical time steps, and $z_{\rm top}$ denotes the vertical coordinate at the top boundary.

In addition to the MHD equations, we also solved the time evolution of the ionization fractions of Oxygen ions as follows.
\begin{eqnarray}
    &&\frac{\partial N_{\rm i}}{\partial t} + \nabla \cdot \left( N_{\rm i} \mathbf{v} \right) \nonumber \\
    &=& N_{\rm e} \left[
    S_{\rm i-1} N_{\rm i-1} + \alpha_{\rm i+1} N_{\rm i+1} 
    - \left( S_{\rm i} + \alpha_{\rm i} \right) N_{\rm i}
    \right],
\end{eqnarray}
where $N_{\rm i}$ denotes the number density of the ions in ith stage, while $N_{\rm e}$ indicates the electron number density.
The coefficients on the right-hand side, $S_{\rm i}$ and $\alpha_{\rm i}$, indicate temperature-dependent collisional ionizaion and recombination rate coefficients obtained from CHIANTI atomic database 10.0.2.
These rates are provided in the low-density limit; however, incorporating density-dependent corrections has been shown to reduce discrepancies between theoretical predictions and observations for Li- and Na-like ions \citep{1979ApJ...228L..89V,1981ApJ...245.1141R,2023MNRAS.521.4696D,2024ApJ...971...59B}.
The feedback of ionization population to the radiative loss function is not considered for simplicity.
{However, such feedback effects can enhance the radiative losses by up to a factor of four compared to equilibrium predictions \citep{1982ApJ...255..783M,1993ApJ...402..741H} and can modify the plasma dynamics in response to nanoflares \citep{2003A&A...407.1127B}.
The feedback effects will be explored in the future study.}
We applied the upwind scheme with a second-order van Leer limiter to track the time evolution of the ionization fractions.

The two-dimensional simulation extends from the upper convection zone, 2 Mm below the photosphere, to the lower corona, 14 Mm above the photosphere, with a horizontal span of 3.84 Mm.
Periodic boundary conditions are applied horizontally. 
The bottom boundary is set as an open boundary with a vertical magnetic field, consistent with the $OV$ configuration described in \cite{2014ApJ...789..132R}. 
The top boundary is treated as a reflecting boundary.

AMR techniques were employed in our simulations, achieving a {grid spacing} of 1.25 km near the TR. 
The AMR routines were adapted from the \texttt{Athena++} code \citep{2020ApJS..249....4S}. 
In this framework, the entire computational domain is divided into \textit{MeshBlocks}, each containing $n \times m$ cells in 2D case ($n=m=16$ in this study). 
When a \textit{MeshBlock} satisfies a refinement criterion (${\rm max}(|\nabla T|/T/\Delta) > 0.1 $), where $\Delta$ is the local grid size, it is subdivided into four child \textit{MeshBlocks} with double the spatial resolution. 
Conversely, refined \textit{MeshBlocks} are de-refined if they meet a de-refinement criterion (${\rm max}(|\nabla T|/T/\Delta) < 0.05$).
The root \textit{MeshBlocks} have a grid size of 40 km in both the horizontal and vertical directions, allowing for a resolution of 1.25 km with six levels of refinement. 
The maximum refinement level is restricted to three (10 km resolution) in regions where $z \leq 1$ Mm or $T \leq 10,000$ K, to prevent unnecessary refinement in chromospheric shocks.

The quasi-steady state was achieved through the following procedure. We began with a one-dimensional standard solar model \citep[e.g.,][]{2021LRSP...18....2C} without a magnetic field as the initial condition, allowing the system to evolve for 5 hours to obtain quasi-steady convective motions with a {grid spacing} of 40 km. 
Subsequently, a uniform vertical magnetic field of 5 G was introduced, and the simulation was allowed to evolve for an additional hour. 
Following this, we started to refine/de-refine \textit{MeshBlocks}, and the simulation was continued for another 2,780 seconds. 
The final 20 minutes of the simulation were used for analysis.

Finally, we conducted 1D simulations to investigate the effects of multi-dimensional processes. 
In the absence of surface convection in the 1D case, we introduced an artificial periodic driving force near the photosphere with a period of 180 seconds. 
The finest {grid spacing} in the 1D simulation was set to 4.9 meters, ensuring that the temperature difference between adjacent grid points, normalized by the local temperature, remained below 0.1 throughout the domain. 
All other simulation parameters were kept identical to those used in the 2D simulations.

\section{Results and Discussions} \label{sec:result}

\subsection{Basic properties of the model}

\begin{figure*}
	\includegraphics[width=\linewidth]{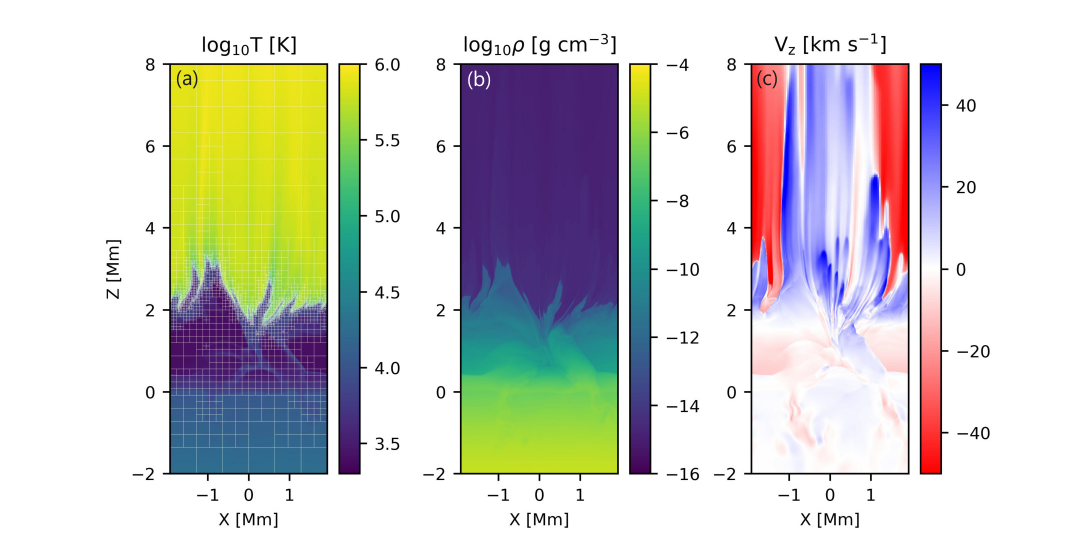}
    \caption{ Temperature (a), mass density (b), and vertical velocity (c) maps at $t$ = 24,310 sec are shown. 
    The white boxes in the panel (a) indicates \textit{MeshBlocks} that contain 16$\times$16 cells.
    }
    \label{fig:2dmaps}
\end{figure*}

We analyzed the model in a quasi-steady state, where the TR moves up and down due to interactions with magneto-acoustic shocks generated by the turbulent convection.
These motions of the TR are typically interpreted as the spicular motions \citep{1982SoPh...75...35H, 1982SoPh...75...99S, 1999ApJ...514..493K} and have been extensively studied by several authors using multi-dimensional radiative MHD simulations \citep[e.g.][]{2006ApJ...647L..73H,2007ApJ...655..624D,2009ApJ...701.1569M, 2015ApJ...812L..30I}.
The snapshots of temperature, mass density, and vertical velocity are shown in Figure \ref{fig:2dmaps}.
The white boxes in panel (a) indicates \textit{MeshBlocks} that contains 16$\times$16 cells.
The finest \textit{Meshblocks} are allocated {near the TR where the temperature gradient is large.}

To examine the resolution dependence of the fundamental properties of our model, we conducted six simulations with varying maximum AMR levels (lvmax), ranging from 1 to 6, corresponding to minimum grid sizes from 40 km to 1.25 km.

\begin{figure}
	\includegraphics[width=9cm]{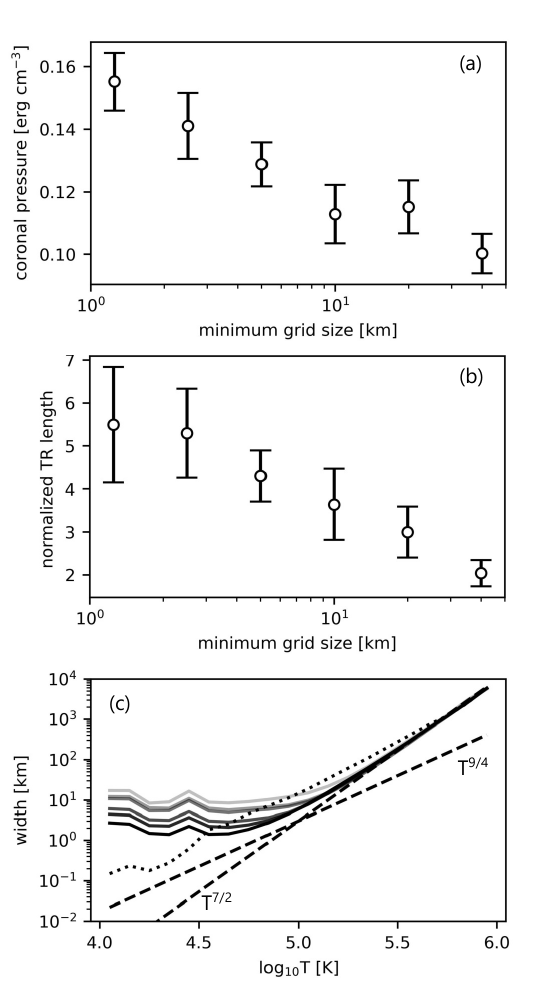}
    \caption{ Resolution dependence of (a) coronal pressure, (b) normalized length of the TR, and (c) effective width of the TR.
    The solid lines in panel (c) represent the results from lvmax=1 to lvmax=6, with lighter colors corresponding to lower levels. The two dashed lines are overplotted for reference, representing proportionality to $T^{9/4}$ and $T^{7/2}$, respectively.
    The dotted line represents the result from 1D simulation.
    }
    \label{fig:resolution_dependence}
\end{figure}

The coronal pressure has not fully converged at the current resolution and continues to increase as the finest grid size is reduced to 1.25 km. 
Figure \ref{fig:resolution_dependence}a shows the coronal pressure, averaged over regions where the temperature exceeds 10$^5$ K, as a function of the minimum grid size. 
{The vertical bars correspond to the standard deviation over a 20-minute period.}
A possible explanation for the pressure increase is enhanced coronal heating due to the increased energy flux into the corona \citep{2016MNRAS.463..502M}.
Furthermore, as discussed later, the lower TR ($\log_{10} T \lesssim 5$) remains under-resolved even at this resolution, potentially contributing to density increases with grid refinement \citep{2013ApJ...770...12B}, despite the use of the LTRAC method.

The length of the TR, representing its corrugation, has marginally converged at the current resolution. 
Figure \ref{fig:resolution_dependence}b shows the TR length (area in 3D case), calculated as the length of the 10$^5$ K temperature contours, normalized by the horizontal extent of the simulation (3.84 Mm in this study), as a function of the minimum grid size. 
As the resolution increases, the corrugation of the TR develops finer structures, resulting in an increase in its length. 
In the present model, the length of the TR is nearly 5.5 times greater than its corresponding horizontal extent.
Assuming convergence at a 2.5 km grid size and requiring approximately 10 grid points to fully resolve the structure, we estimate a minimum spicule width of around 25 km, potentially below the spatial resolution (0.1-0.2\arcsec) of recent observational studies on spicule widths \citep[e.g.][]{2012ApJ...759...18P}.
Additionally, these corrugations, or spicules, increase the (E)UV emissions when viewed at the solar limb \citep{1978ApJ...226..698M}.
However, the increase in TR length does not imply an increase in TR volume, as the TR width remains nearly constant, not perpendicular to the TR but rather along the magnetic field direction.

The upper TR ($\log_{10} T \gtrsim 5$) is likely fully resolved at the current resolution; however, finer spatial resolution is still required to adequately resolve the lower TR. 
Figure \ref{fig:resolution_dependence}c shows the effective width of the TR as a function of temperature. 
The effective width is calculated as the volume of gas within a specific temperature range, normalized by the horizontal extent of the simulation. 
The solid lines represent results from maximum AMR levels of 1 to 6, with lighter colors corresponding to lower levels. 
The two dashed lines are overplotted for reference, representing proportionality to T$^{9/4}$ and T$^{7/2}$, respectively.
{The proportionality for the lower TR is derived from the Field length $L$ \citep{1990ApJ...358..375B}, which follows from energy balance equation, $T^{7/2}/L^2 \propto n^2 T$, assuming constant pressure.
This relationship is widely applied to estimate the interface width between the cold and warm interstellar medium and can similarly be extended to characterize the width of the TR.
In contrast, the proportionality for the upper TR arises from the condition of constant thermal conduction flux, expressed as $T^{7/2}/L$ = constant.}
The dotted line represents results from a 1D simulation with a minimum grid size of 4.9 m for comparison. 
Although the estimated width shows marginal convergence down to log$_{10}T\sim$4.8, finer {grid spacing} is still required for a fully resolved lower TR.
Resolving the upper TR is crucial for accurately solving the advection equations for the ionization fractions of species like \ion{O}{4} or higher ionization degree, which form at temperatures exceeding 10$^5$ K.
The results for species like \ion{O}{3} or lower ionization degree will probably change as we use finer grid size.
The stringent spatial resolution requirement for the lower TR may be alleviated if the effective cooling rate is reduced through mechanisms such as backward coronal radiation \citep{2012A&A...539A..39C} or if thermal conduction is suppressed in partially ionized plasma \citep{1968JPlPh...2..617D}.
{
Because the grid spacing in our simulations is typically smaller than or comparable to the electron mean free path in the corona and TR, non-MHD effects must be considered \citep{2015RSPTA.37350055P}. These effects can modify the width of TR by changing the classical Spitzer thermal conductivity. Addressing this issue fully may require a detailed examination of the velocity distribution functions, similar to approaches used in studies of the solar wind \citep{1982JGR....87.5030M}. However, such analysis lies beyond the scope of the present study.
}

\begin{figure}
	\includegraphics[width=9cm]{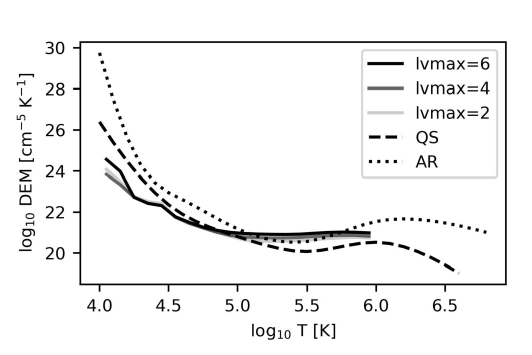}
    \caption{ Resolution dependence of DEM. 
    The solid lines represent the results from maximum level = 2, 4 and 6, with lighter colors corresponding to lower levels.
    The dotted and dashed line indicate the DEM derived using the data from \citep{1978ApJS...37..485V} for quiet Sun and active region, assuming coronal abundance and a constant pressure of 3$\times$10$^{15}$ cm$^{-3}$ K, respectively.
    }
    \label{fig:dem}
\end{figure}

Further adjustments are required to reproduce the differential emission measure (DEM) in agreement with observations. 
Figure \ref{fig:dem} shows the DEM as a function of temperature. While the DEM at log$_{10}$T $\sim$ 5 is consistent with observations \citep{1978ApJS...37..485V}, the lower temperature range is significantly underestimated compared to observations, and the higher temperature range is slightly overestimated. 
Additionally, the DEM in the higher temperature range increases with spatial resolution, likely reflecting the resolution dependence of density (or pressure). 
The well-known discrepancies in the lower TR DEM between models and observations have been linked to contributions from low-lying loops \citep{1986SoPh..105...35D}. 
Recent numerical studies further suggest that emission from \ion{Si}{4} 1393 {\AA}, forming around 5-8 $\times 10^4$ K , is predominantly contributed by short loops \citep{2016ApJ...831..158S}.

\subsection{Properties of non equilibrium ionization}

\begin{figure*}
    \includegraphics[width=\linewidth]{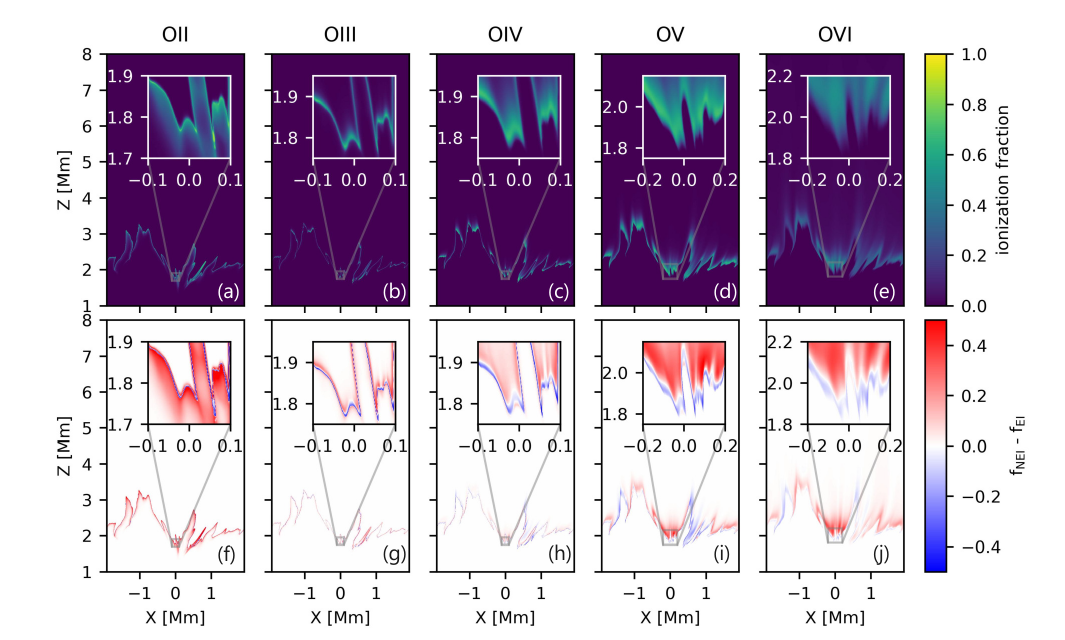}
    \caption{ (a–e) Two-dimensional maps of the ionization fractions of \ion{O}{2} to \ion{O}{6}, and (f–j) two-dimensional maps of the difference in ionization fractions between NEI and EI. Each panel includes a zoomed-in view of a sub-region of the maps.
    The time of the snapshot is identical to that of Figure \ref{fig:2dmaps}.
    }
    \label{fig:2d_ion_frac}
\end{figure*}

The spatial distribution of ionization fractions shows significant deviations from those in equilibrium ionization. 
Figure \ref{fig:2d_ion_frac} presents two-dimensional maps of the ionization fractions of \ion{O}{2} to \ion{O}{6} (a–e), as well as the difference in ionization fractions between NEI and equilibrium ionization (EI) (f–j). 
The ionization fraction peaks for \ion{O}{4} to \ion{O}{6} are well-resolved, whereas those for \ion{O}{2} and \ion{O}{3} are very sharp and remain under-resolved at the current spatial resolution. 
The snapshot (t = 24,310 sec) corresponds to the same time as in Figure \ref{fig:2dmaps}, occurring just after the chromospheric shock interacts with the TR, as indicated by the positive v$_{\rm z}$ immediately above the TR (Figure \ref{fig:2dmaps}c). 
During this period, chromospheric material evaporates into the corona, resulting in a clear transition from under-abundance at lower heights to over-abundance at higher heights for \ion{O}{4} to \ion{O}{6}. 
As we will demonstrate later, the opposite spatial distribution is observed during the condensation phase.

\begin{figure}
	\includegraphics[width=9cm]{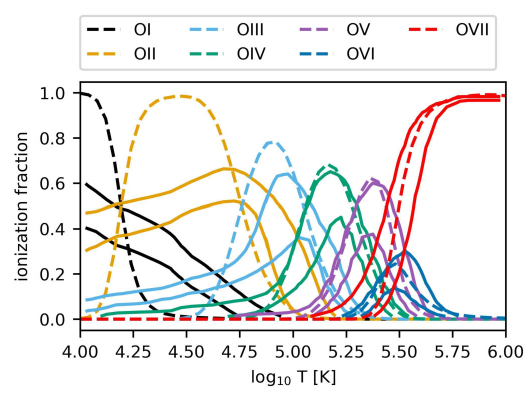}
    \caption{
        The probability distribution of the ionization fractions of oxygen ions as a function of temperature. 
        The solid lines represent the one-sigma intervals for NEI fractions, while the dashed line indicates the fractions under EI conditions.
    }
    \label{fig:pdf_ionfrac}
\end{figure}

The aforementioned differences in the ionization fractions of oxygen ions are also evident in the probability distribution functions (PDFs) sampled during the final 1,200 seconds of the simulation (Figure \ref{fig:pdf_ionfrac}). 
The PDFs, denoted as $F_{\rm i}$, are computed as
\begin{eqnarray}
    F_{\rm i}(T;x) = P(T; f_{\rm i} \le x),\\
    \int_0^1 F_{\rm i}(T;x) dx = 1
\end{eqnarray}
where i serves as the index of ion species, and $P(T;f_{\rm i} \le x)$ represents the probability that the ionization fraction $f_{\rm i}$ is smaller than $x$ at temperature $T$. 
The PDFs for \ion{O}{4} to \ion{O}{6} show slight under-abundances, although each PDF exhibits a minor over-abundance at the higher temperature range. 
For \ion{O}{1} to \ion{O}{3}, pronounced deviations between NEI and EI are observed, which can be attributed to the shorter cooling timescales relative to ionization timescales in the lower TR.
The broad tail structure in the ionization fraction of \ion{O}{2} (Figure \ref{fig:2d_ion_frac}a, f) may suggest that plasma undergoing rapid cooling is recombining to lower ionization states. 
This deviation may also be partially due to numerical diffusion effects resulting from insufficient spatial resolution, especially for \ion{O}{3} that have sharp spatial distribution (Figure \ref{fig:2d_ion_frac}b, g).

%We attribute the significant deviations between NEI and EI for \ion{O}{1} to \ion{O}{3} to numerical diffusion in the advection equations due to insufficient spatial resolution, as seen in Figure \ref{fig:resolution_dependence}c and Figure \ref{fig:2d_ion_frac}.

To assess the model, we estimate intensities emitted from ions in NEI ($I_{\rm NEI}$) and EI ($I_{\rm EI}$), the Doppler velocity, and nonthermal width, which are fundamental properties observable in solar data. 
As a representative example, resolution dependence of these quantities were calculated for the emission lines at 1031.91 {\AA} from Li-like \ion{O}{6} ion, where the ionization fraction is spatially well-resolved. 

\begin{figure}
    \includegraphics[width=9cm]{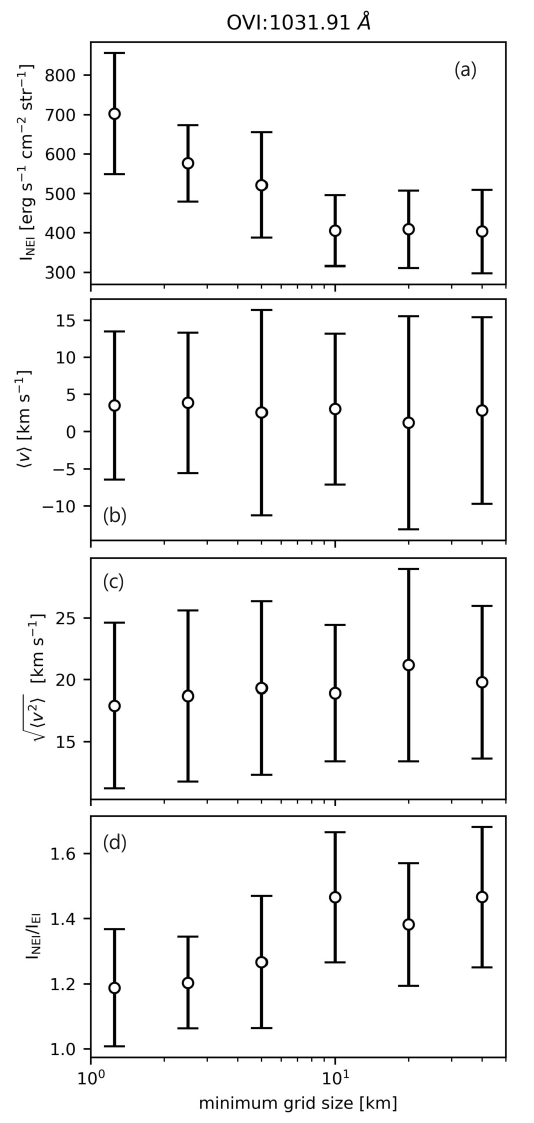}
    \caption{ 
    Resolution dependence of (a) intensity, (b) Doppler velocity, (c) nonthermal width, and (d) the ratio of $I_{\rm NEI}$ to $I_{\rm EI}$ for the \ion{O}{6} emission line at 1031.91 \AA.
    }
    \label{fig:resolution_dependence_of_line}
\end{figure}

The TR line intensity at 1031.91 \AA\ increases with higher spatial resolution (Figure \ref{fig:resolution_dependence_of_line}a), likely reflecting the resolution dependence of coronal pressure. 
These intensities are consistent with quiet Sun observations \citep{1995ApJ...455L..85J,2023MNRAS.521.4696D}.

The Doppler velocity of the TR line at 1031.91 \AA\ has fully converged even at the lowest resolution, showing no significant average shifts (Figure \ref{fig:resolution_dependence_of_line}b). 
This result is inconsistent with observations, which typically show redshifts in TR lines both in the quiet Sun \citep{1998ApJS..114..151C, 1999ApJ...522.1148P} and in equatorial coronal holes \citep{2004A&A...424.1025X}. 
Addressing this discrepancy will require 3D models that account for episodic heating due to magnetic field braiding \citep{2006ApJ...638.1086P, 2010ApJ...718.1070H, 2022A&A...661A..94C}.

The nonthermal velocity has also converged to approximately 20 km s$^{-1}$ at the lowest resolution (Figure \ref{fig:resolution_dependence_of_line}c), although this value is smaller than observed values of around 30 km s$^{-1}$ \citep{1998ApJ...505..957C, 2001A&A...374.1108P}. 
A similar discrepancy in nonthermal width in the middle TR has also been reported in 3D simulations \citep{2006ApJ...638.1086P}, suggesting that higher spatial resolution may be required to resolve small-scale velocities induced by nanoflares.
{
To accurately reproduce the observed nonthermal widths, it will be essential to extend the current model to three dimensions and ensure sufficient spatial resolution to capture the nanoflare processes.
}

The ratio of $I_{\rm NEI}$ to $I_{\rm EI}$ has converged to approximately 1.2 at the highest resolution (Figure \ref{fig:resolution_dependence_of_line}d). 
Lower resolutions tend to exhibit higher ratios, likely due to numerical diffusion in the advection equations caused by insufficient spatial resolution. 
Observationally, the ratio between the measured and theoretical intensities (assuming EI) of Li-like oxygen ion, \ion{O}{6}, is generally around 2 in the quiet sun \citep{1995ApJ...455L..85J, 2023MNRAS.521.4696D}. 
Accounting for NEI effects, the theoretical intensity, $I_{\rm NEI}$, increases by 20\%, reducing the observed-to-theoretical intensity ratio to approximately 1.7. 
While NEI effects mitigate some of the anomalous behavior observed in Li-like ions, additional mechanisms are required to fully explain this discrepancy.

\begin{figure}
    \includegraphics[width=9cm]{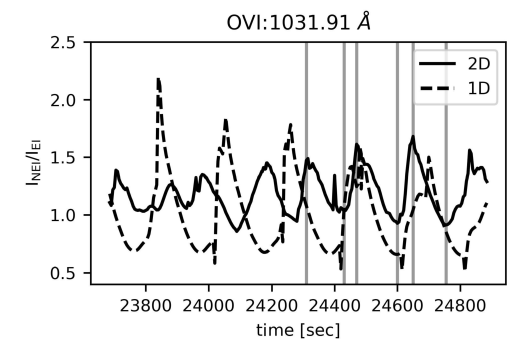}
    \caption{ 
    The time evolution of the ratio between $I_{\rm NEI}$ and $I_{\rm EI}$. 
    Solid lines represent the results from 2D simulations, while dashed lines correspond to the 1D simulations. 
    Note that although the duration is the same for both 2D and 1D simulations, the phases are independent of each other.
    }
    \label{fig:ratio_evolution}
\end{figure}

To further examine the behavior of $I_{\rm NEI}/I_{\rm EI}$, its time evolution is shown in Figure \ref{fig:ratio_evolution}, revealing periodic oscillations with a period of approximately 3 minutes. 
Solid lines represent the results from 2D simulations, while dashed lines correspond to 1D simulations. 
These oscillations arise from the periodic collisions between chromospheric shocks and the TR. 
The amplitude of the oscillations is smaller in the 2D simulations compared to the 1D case, likely due to superposition effects inherent in the 2D model.
In contrast to the 2D case, the average value of $I_{\rm NEI}/I_{\rm EI}$ in the 1D case does not exhibit a significant deviation from 1, which is consistent with previous study \citep{2024ApJ...964..107M}.

\begin{figure*}
	\includegraphics[width=\linewidth]{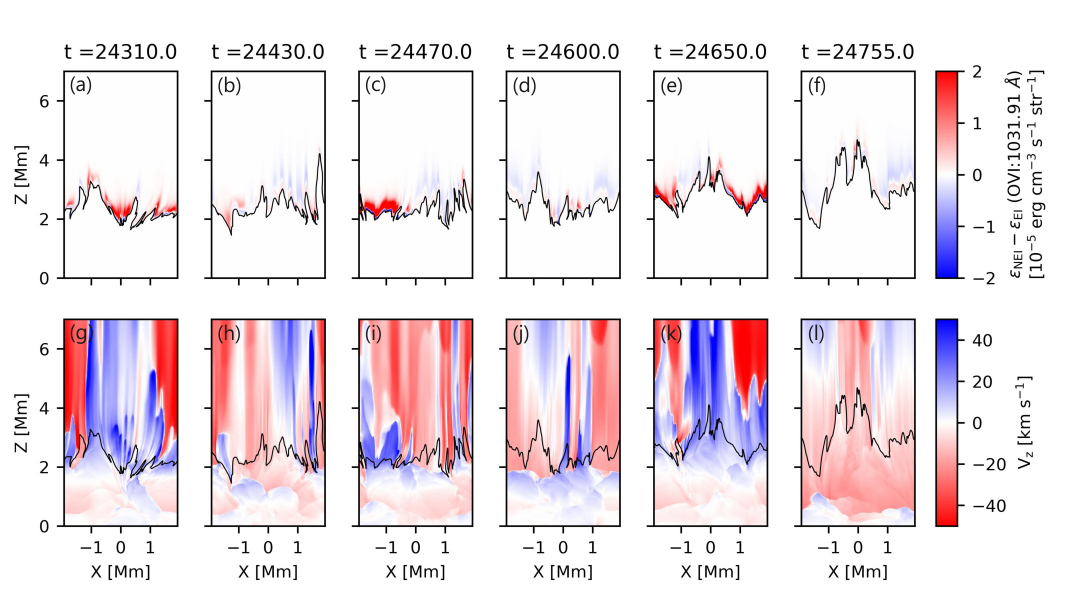}
    \caption{ 
    Difference maps between $\epsilon_{\rm NEI}$ and $\epsilon_{\rm EI}$ for \ion{O}{6} ions (1031.91 \AA, upper panels), along with vertical velocity maps (lower panels), are shown at multiple time points. The selected times correspond to the peaks (a, c, e, g, i, k) and troughs (b, d, f, h, j, l) of the plot in Figure \ref{fig:ratio_evolution}, as indicated by the gray vertical lines. The black contours in each panel delineate the TR, where $T = 10^5$ K.
    }
    \label{fig:shock_tr_collision}
\end{figure*}

The collision between chromospheric shocks and the TR enhances the emission from \ion{O}{6} ions beyond what is predicted by the EI approximation, while the return flow reduces the emission (Figure \ref{fig:shock_tr_collision}). 
The upper panels display 2D maps showing the difference between $\epsilon_{\rm NEI}$ and $\epsilon_{\rm EI}$ for \ion{O}{6} ions (1031.91 \AA) at various times, where $\epsilon_{\rm NEI}$ and $\epsilon_{\rm EI}$ represent the emission from ions under NEI and EI conditions, respectively. 
The selected times correspond to the peaks and troughs of the time evolution of $I_{\rm NEI}/I_{\rm EI}$, indicated by gray vertical lines in Figure \ref{fig:ratio_evolution}.
In the odd column of the figure (Figure \ref{fig:shock_tr_collision}a, c, e, g, i, k), strong upward flows above the TR are observed, originating from chromospheric shocks. 
The region beneath these upflows and just above the $T = 10^5$ K contour (the black solid curves in each panel) shows positive $\epsilon_{\rm NEI} -\epsilon_{\rm EI}$. 
The abrupt temperature increase in this region due to the shock passages likely causes deviations from equilibrium, leading to enhanced emission. 
Conversely, in the even column of the figure (Figure\ref{fig:shock_tr_collision}b, d, f, h, j, l), downward flows are dominant. 
In these cases, the region beneath the downflows and above the $T = 10^5$ K contour shows negative $\epsilon_{\rm NEI} -\epsilon_{\rm EI}$.
The gradual temperature decrease likely causes deviations from equilibrium, resulting in reduced emission.

\section{Conclusions} \label{sec:concl}

We conducted two-dimensional radiative MHD simulations from the surface convection zone to the corona. 
By implementing AMR, our model resolves the upper TR (log$_{10} T \gtrsim 5$), capturing the time evolution of oxygen ionization fractions in this region with adequate spatial resolution. 
However, the lower TR (log$_{10} T \lesssim 5$) remains under-resolved even with a minimum grid size of 1.25 km. 
Reducing the minimum grid size from 40 km to 1.25 km led to increases of 1.5-fold in coronal pressure and 5.5-fold in TR length. 
While the mean intensity aligns with observations of the quiet Sun, the Doppler shifts and nonthermal widths of the synthesized emission lines do not fully agree with observational data. 
Nevertheless, the anomalous enhancement of Li-like oxygen ions, specifically \ion{O}{6}, in our model narrows the gap between observations and theoretical predictions under equilibrium ionization conditions, though additional enhancement would be necessary to fully account for the observed values.

Spatially resolving the TR is essential to accurately reproduce emissions from ions in this region, especially those in non-equilibrium ionization states. 
Investigating these emissions will be critical for ongoing and upcoming space missions, such as Solar Orbiter/SPICE \citep{2013SPIE.8862E..0FF} and the future Solar-C/EUVST mission \citep{2020SPIE11444E..0NS}, both of which will capture a broad UV spectral range. 
These missions offer unprecedented opportunities to compare observed and modeled TR emissions, potentially refining our understanding of mass and energy transport in the solar atmosphere.

\begin{acknowledgments}
We extend our heartfelt gratitude to the reviewer for their insightful comments and constructive feedback, which greatly enhanced the quality of this manuscript.
This work was supported by JSPS KAKENHI Grant Number JP23K03456.
2D numerical computations were carried out on Cray XC50 at the Center for Computational Astrophysics, National Astronomical Observatory of Japan.
Data analysis and 1D simulations were carried by using the computational resource of the Center for Integrated Data Science, Institute for Space-Earth Environmental Research, Nagoya University through the joint research program.
CHIANTI is a collaborative project involving George Mason University, the University of Michigan (USA), University of Cambridge (UK) and NASA Goddard Space Flight Center (USA).
\end{acknowledgments}

\bibliography{myref}{}
\bibliographystyle{aasjournal}

\end{document}